\documentclass[a4paper,11pt]{scrartcl}

\usepackage{fullpage}
\usepackage{lineno,hyperref}

\usepackage{graphicx}
\usepackage{amsmath}
\usepackage{bm}

\usepackage{color}

\begin{document}

\title{Optimizing Scrubbing by Netlist Analysis for FPGA Configuration Bit Classification and Floorplanning\thanks{This article has been published in the Journal \emph{Integration, the VLSI Journal} 59C (2017) pp. 98-108, \newline DOI: \texttt{10.1016/j.vlsi.2017.06.012}\newline \textcopyright 2017. This manuscript version is made available under the CC-BY-NC-ND 4.0 license\newline \url{http://creativecommons.org/licenses/by-nc-nd/4.0/}}}
\author{Bernhard Schmidt, Daniel Ziener, J\"urgen Teich, and Christian Z\"ollner \\[2mm] \large
Hardware/Software Co-Design, Department of Computer Science,\\ \large
Friedrich-Alexander-Universit\"at Erlangen-N\"urnberg (FAU)\\ \large
Cauerstrasse 11, 91058 Erlangen, Germany\\ \large
Email: \{daniel.ziener, juergen.teich\}@fau.de}
\date{ }
\maketitle

\begin{abstract}
Existing scrubbing techniques for SEU mitigation on FPGAs do not guarantee an error-free operation after SEU recovering if the affected configuration bits do belong to feedback loops of the implemented circuits.
In this paper, we a) provide a netlist-based circuit analysis technique to distinguish so-called \emph{critical} configuration bits from \emph{essential} bits in order to identify configuration bits which will need also state-restoring actions after a recovered SEU and which not. Furthermore, b) an alternative classification approach using fault injection is developed in order to compare both classification techniques. Moreover, c) we will propose a floorplanning approach for reducing the effective number of scrubbed frames and d), experimental results will give evidence that our optimization methodology not only allows to detect errors earlier but also to minimize the Mean-Time-To-Repair (MTTR) of a circuit considerably.
In particular, we show that by using our approach, the MTTR for datapath-intensive circuits can be reduced by up to 48.5 \% in comparison to standard approaches.
\end{abstract}


\section{Introduction}

The application areas of Field Programmable Gate Arrays (FPGAs) have steadily grown over the last two decades. Initially, FPGAs were intended to replace CPLDs to implement simple glue logic functions, but today's SRAM-based FPGAs are able to host entire multi-core systems on a single chip. By now, FPGAs play a major role for implementing complex digital systems for telecommunication networks, software-defined radios, and computer vision. The success in these application areas makes FPGAs also interesting for new safety-critical applications and application fields, like in  space missions or avionics, which are still in the hand of traditional ASICs. However, in these new operation sites, FPGAs have to deal with harsh environments, and the implemented systems are forced to guarantee a high reliability.  

Today, most advanced SRAM-based FPGA devices such as the Xilinx Virtex-7 family offer up to 2 million logic cells. On the other hand, these huge numbers of FPGA resources must be configured which results in bitstream sizes of up to 55 MB \cite{DS180}. In fact, this makes current FPGA devices to some of the largest SRAM chips available \cite{dagstuhl10koch}. 
This development was only possible by advances in manufacturing process technology to ever-smaller scales. In general, the SRAM-cells of an SRAM chip are susceptible to effects of cosmic ray particles, which can lead to so called \emph{Single Event Upsets} (SEUs), a bit-flip of the stored value in one SRAM-cell. Although, the charges stored in SRAM-cells decrease with every new technology generation, the susceptibility to SEU of one SRAM-cell has been lowered or has been held constant at least \cite{XRR13}. Nonetheless, since the number of SRAM-cells per FPGA-device has been increased significantly, the probability that an SEU occurs during operation of the FPGA device has been increased. This makes SEUs an important reliability issue, especially in harsh radiation-prone environments like space applications. Here, SEUs impose a well-established concern, but they receive increasing concern also for safety-critical terrestrial applications such as systems for medical, automotive, and power generation 
applications.

To prevent the accumulation of SEUs, well-known frame-based scrubbing techniques \cite{XAPP216} can be successfully applied. Scrubbing allows the recovery from errors induced by SEUs by periodically or error-triggered frame-wise overwriting of configuration bits with the unfalsified values. 

However, most of these approaches do not take into account that SEUs have varying impacts on the implemented logic. Even in large designs, only a minority of the configuration bits is used and, therefore, have influence on the implemented functionality. SEUs on \emph{unused} configuration bits may obviously be ignored. However, SEUs on unused configuration bits might create some ghost circuits which might increase the FPGA power consumption, but have no influence on the functionality of the design.  Moreover, SEUs on used configuration bits may also have varying impacts on the circuit behavior and may demand different error-handling methods. In particular, not all SEU-induced errors may be corrected only by the application of scrubbing. Indeed, components of a circuit involved in feedback paths might still and infinitely cause a malfunction due to corrupted states even if scrubbing is applied to the involved configuration frames. 
 
\begin{figure}[t]
\centering
\includegraphics[width=3.6in]{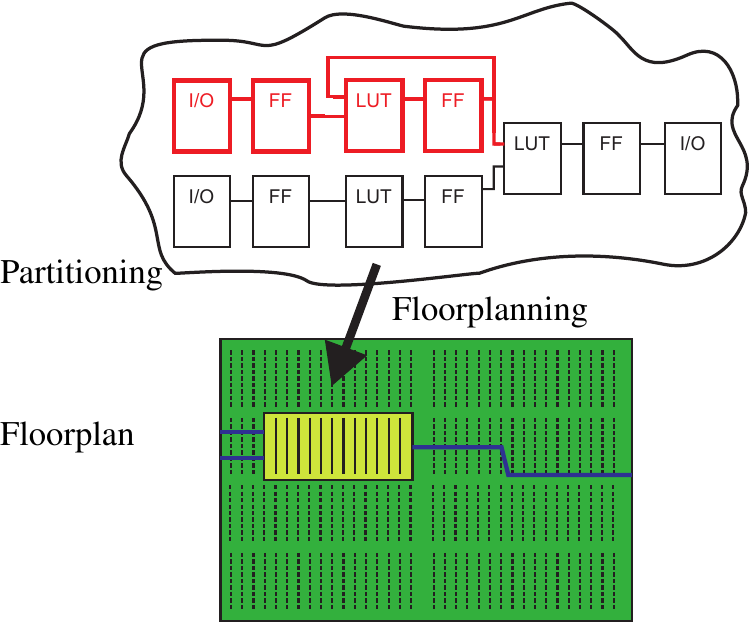}
\caption{Illustration of our two step approach. In the partitioning step, the primitive cells of the netlist, e.g., LUTs and flip-flops, and nets are categorized into \emph{essential} (black) and \emph{critical} (red) cells, nets respectively in order to identify and distinguish the associated \emph{essential} and \emph{critical} bits. In the floorplanning step, the primitive cells are placed and routed such to minimize the number of occupied configuration frames by using of special placement and routing constraints.}
\label{intropic}
\end{figure}

In this paper, we present an automatic partitioning approach which categorizes primitive cells and nets at logic level into \emph{essential} and \emph{critical} cells, respectively nets. Functional errors caused by SEUs on configuration bits of \emph{essential}, but \emph{non-critical} primitive cells and nets may be corrected by scrubbing without any further actions. 
For SEUs on configuration bits of \emph{critical} primitive cells and nets, further actions, like resetting the registers, must be performed additionally to scrubbing. Some related work using the terms \emph{sensitive} and \emph{persistent} instead of \emph{essential} and \emph{critical}, e.g., \cite{Keith05}.
In addition to this automatic netlist partitioning as illustrated in Fig. \ref{intropic}, we propose advanced floorplanning methods to reduce the overall number of frames which have to be scrubbed. As an important side effect, this also may lead to a reduction of the Mean-Time-To-Repair (MTTR) of a given system due to shorter scrubbing cycles. Furthermore, by knowing the classification of each configuration bit, the number of time-consuming state-restoring actions, like  global resets of the circuit, are reduced, since not every SEU will demand such actions after scrubbing.  A set of experiments will give evidence that the proposed technique may indeed reduce the downtime of a system considerably. Furthermore, an alternative classification approach by utilizing fault injection experiments is presented and a comparison to the introduced automatic partitioning approach is provided.

Circuits for the control path consist in general of more feedback loops compared to datapath-intensive circuits. In harsh environments, like space applications, the control-intensive part is mostly implemented into radiation hardened external microprocessors. Therefore, our focus is on FPGA-based datapath-intensive circuits which are common in digital signal processing applications.

This paper extends a preliminary publication \cite{szt14} in the following major aspects: Apart from the presentation of an automated netlist classification and floorplanning method, the design flow in Section \ref{sec:goal} and an alternative fault injection classification approach in Section \ref{sec:faultinj} are introduced, followed by a comparative study of the results of both classification techniques in Section \ref{sec:analysis}. 

The remainder of this paper is structured as follows: Section~\ref{sec:relatedwork} discusses related work in the field. The problem statement and target applications are described in Section \ref{sec:problem}. In Section \ref{sec:notation} and Section \ref{sec:notation22}, we provide formal definitions to describe the correlation between netlist primitive cells, respectively nets and the corresponding configuration bits. Furthermore, we introduce our definition of the MTTR. Our methodology is described in detail in Section \ref{sec:background}, whereas the design flow to implement our efficient SEU scrubbing method is presented in Section \ref{sec:goal}. Section \ref{sec:faultinj} describes the implementation of an alternative classification approach by using fault injection. In Section \ref{sec:analysis}, we present experimental results and Section \ref{sec:conclusion} concludes the paper.  
\vspace*{-0.15cm}

\section{Related Work}
\label{sec:relatedwork}

Previous research has studied the impact of SEUs on SRAM FPGA devices\footnote{\scriptsize{The configuration memory of flash-based FPGAs like the FPGAs of the Microsemi ProAsic3 family is immune to SEUs. However, in general, flash-based FPGAs have a significant lower logic density in comparison to SRAM-based FPGAs \cite{PROASIC13}.}} \cite{bernard11,jing11,legat10}. Many techniques have been proposed to provide highly reliable FPGA devices, e.g., \emph{rad\-iation-hardened FPGAs}\footnote{\scriptsize{Examples are Xilinx Virtex-4QV \cite{DS653}, Xilinx Virtex-5QV \cite{DS692}, Microsemi RTG4 \cite{microredhard}, and the NanoXplore NXT-32000 \cite{nanoexplore}.}}, to lower the effect of rad\-ia\-tion-induced SEUs. However, rad\-ia\-tion-hardened SRAM FPGAs typically have a lower density compared to non rad\-iation-hardened FPGAs, and they only lower the probability of SEUs and do not completely avoid them. A comparison between a rad\-iation-hardened FPGA (Virtex-5QV) and a standard FPGA (Kintex-7) was done in \cite{ahs15}. Even on radiation-hardened FPGAs, the SEU rate in a low-earth orbit can be up to 16  events per day for a Virtex-4QV \cite{quinn12} and up to 1.2 events per day for a Virtex-5QV \cite{fccm14}.  
Hence in space missions, SEU correction mechanisms become essential to avoid the accumulation of latent faults and ensure correct operation of an FPGA.

A wide variety of other SEU fault mitigation techniques for SRAM-based FPGAs have been proposed during the past years. These techniques can be categorized into module redundancy techniques, e.g., \emph{Dual Module Redundancy} (DMR), \emph{Triple Module Redundancy} (TMR) \cite{Lyons62, sv05, DBLP:conf/fpt/AngermeierZGT11}, \emph{N Module Redundancy} (NMR), and techniques that use \emph{scrubbing} of the FPGA configuration memory \cite{hsw09, XSEM12, SariP11}. Also, the combination of both techniques has been shown to be able to increase the reliability of the FPGA modules significantly \cite{OCG09}. FPGA-based TMR approaches replicate a given module which shall be protected either statically \cite{sv05} or dynamically \cite{DBLP:conf/fpt/AngermeierZGT11}. However, TMR techniques are known to often cause an excessive and unacceptable overhead in terms of power consumption and area. Since the intensity of cosmic rays is not constant but may vary over several magnitudes depending on the solar activity, a worst-case radiation protection is far too expensive in most cases. A self-adaptive system is proposed in \cite{fccm14}, which monitors the current SEU rate and exploits the opportunity of partial reconfiguration of FPGAs to implement redundancy such as TMR on demand. Furthermore, the possibility to partially reconfigure the FPGA could also be exploited to repair faulty modules (single TMR instances) through reconfiguration which may be identified by a majority voter \cite{unsw}. Hereby, it is necessary that the modules do not include any feedback paths. Otherwise, costly state recovering actions must be applied. 

Scrubbing techniques can also be categorized into \emph{blind} and \emph{non-blind} scrubbing. Blind scrubbing refreshes the configuration memory in a periodically manner without any error detection. Non-blind scrubbing, however, refreshes the configuration only once an SEU is detected. Commonly, the error detection of non-blind scrubbing is achieved by a frame-by-frame readback with Error Correction Code (ECC) checking \cite{XSEM12}. In addition, a Cyclic Redundancy Check (CRC) over the complete array of frames is used for a fast detection of errors in the configuration memory.
In \cite{Pratt06} and \cite{Pratt08}, an approach called partial TMR is described in which TMR is selectively applied to a given design to keep the overhead small due to the triplication of combinatorial and sequential logic. To identify locations requiring triplication, the authors propose a netlist analysis. In our approach, also a netlist analysis is proposed, however, in order to identify the configuration bits of \emph{critical} primitive instances and \emph{critical} nets rather than places for module triplication. Finally, in \cite{SariP11}, the authors propose to use placement constraints aligned to the frame boundaries to minimize the number of frames with \emph{essential} bits which is similar to our minimization approach. However, they do not actively manipulate the routing. Moreover, the characterization of \emph{essential} and \emph{critical} bits is not considered in their work.

\section{Problem Statement}
\label{sec:problem}

In general, SEUs in the configuration memory are considered as \emph{soft errors}. A bit-flip in the configuration memory does not cause a permanent and non-recoverable defect, since it does not damage the SRAM cell itself, but corrupts its stored data. A typical manifestation of an SEU-related error is an altered LUT function or signal route. A circuit design may be simply corrected by overwriting the corresponding memory cell with the correct value. However, the configuration bits of an FPGA are typically written once at the FPGA boot-up time and are never changed or refreshed during operation. Therefore, without any SEU correction mechanism, the output errors induced by SEUs appear to be permanent. In this context, we do not consider SEUs, which change the state of the user-logic flip-flops directly, since they do not change the configuration of an FPGA.

As described in Section \ref{sec:relatedwork}, SEUs can be corrected by \emph{scrubbing} techniques. However, this takes some detection and correction time, in which the circuit might also produce and store wrong results in registers. If the circuit netlist is free of cycles, no further actions have to be carried out after scrubbing. However, if the circuit contains any feedback cycles, a recovery of the circuit after \emph{scrubbing} can only be achieved by resetting all registers to an initial state or by replacing the register values by values saved at checkpoints during an uncorrupted circuit behavior. Nevertheless, the application has to tolerate such a behavior by invalidating corrupt output data. Streaming and packet-based applications, in which video, audio, or wireless communication data are processed, usually may tolerate such behavior as well as partially corrupted outputs since data may also be corrupted by interference of wireless transmissions and, therefore, error correction schemes must always be present. 

One example is the receiver chain of a communication satellite presented in \cite{ahs15}. A demodulator, including frequency conversion, filtering, and demapping of a Quadrature Phase-Shift Keying (QPSK) modulation is followed by a deinterleaver and a Low-Density-Parity-Check (LDPC). An erroneous output of this receiver chain might be tolerated for a short period of time due to implemented error handling in upper layers of the communication protocol.  Note that control intensive  designs might not tolerate such a behavior. Therefore, our approach focuses on data paths which are implemented in a stream-based manner.   By combining our approach with TMR, also applications can be supported which do not tolerate this limitation. 

One important reliability metric is the \emph{Mean-Time-To-Repair} (MTTR). Consider a system model that tolerates corrupt outputs for short times and rollbacks to former saved checkpoints. In that case, the MTTR is the period of time in which the circuit may deliver corrupt output values. Therefore, a reduction of the MTTR will be of great benefit increasing the overall circuit reliability. 
On the other hand, the MTTR is closely related to the \emph{Mean-Time-To-Failure} (MTTF) since repair actions have to be taken only if a failure occurs.  
According to~\cite{OCG09}, the SEU-related MTTF depends on the SEU rate $\mu_{\text{FPGA}}$ of the FPGA's configuration memory and the probability that an SEU hits an \emph{essential} bit.
The SEU rate $\mu_{\text{FPGA}}$ depends on the FPGA device\footnote{\scriptsize{The main factors of the SEU sensitivity of FPGA devices are the technology parameters (mainly $V_{dd}$ and the structure size), the \emph{die size}, and the \emph{number of configuration bits} \cite{OCG09}.}}  and the environment\footnote{\scriptsize{The most important parameter of the environment is the \emph{particle flux} that is environmental-dependent (ground or different orbits for space applications). 
A tool to estimate the SEU rate $\mu$ of semiconductor devices in \emph{SEUs/device/s} for different orbits is CREME96~\cite{creme}.} The MTTF differences between space and ground conditions for the same design and device can vary in the range of several magnitudes. Therefore, the MTTR reduction is utmost important, especially for space applications with  high SEU rates of up to several SEUs per day, even on radiation-hardened FPGAs \cite{quinn12}}.

\section{Definitions w.r.t. Partitioning and Floorplanning}
\label{sec:notation}

In order to explain our MTTR minimization approach, we first introduce some formal definitions to clarify the dependency between configuration bits and different FPGA resources which is later used to describe our SEU-aware FPGA design flow in detail in Section \ref{sec:goal}.

\begin{figure}[!tb]
\centering
\includegraphics[width=3.2in]{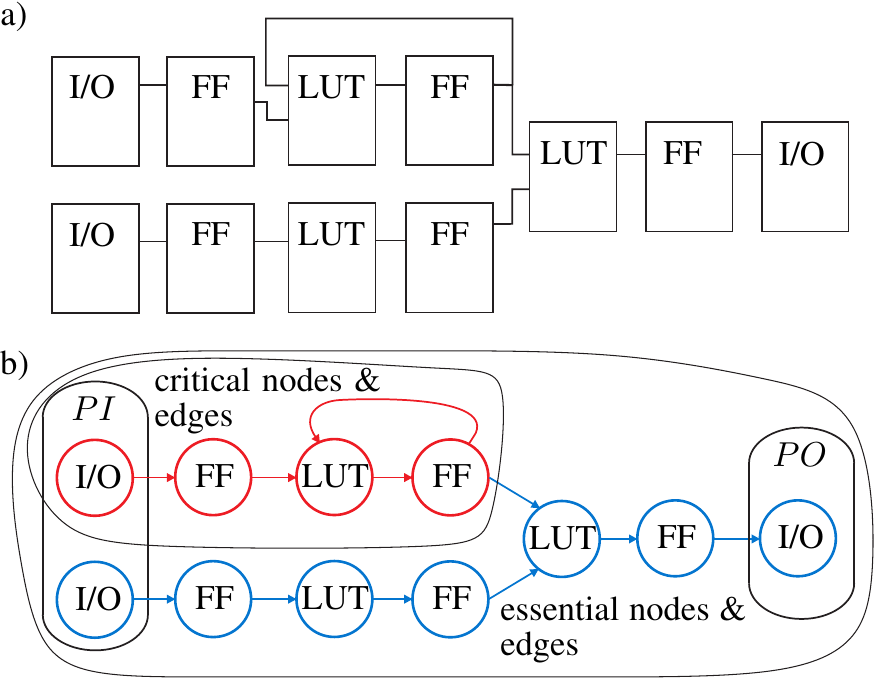}
\caption{Illustration of the logic-level netlist analysis to identify \emph{critical} primitive instances and nets in the set of \emph{essential} primitive instances and nets. The netlist in a) is converted to the directed graph $G(E,V)$ in b) with the primary input nodes $PI$ and the primary output nodes $PO$. \emph{Critical} primitive instances and nets are marked red. \emph{Critical} primitive instances reside on either cycles or on feed-forward paths into cycles and each net which is fed into a \emph{critical} primitive instance is itself \emph{critical} \cite{szt14}.}
\label{example1b}
\end{figure}

During automatic partitioning, we analyze the logic-level netlist of a given digital circuit after logic synthesis. The resulting logic-level netlist is usually given in a structured text format and consists of instances of primitive cells, e.g., \emph{LUTs}, \emph{flip-flops}, \emph{I/Os}, \emph{block memories} (BRAM), \emph{embedded multipliers} and \emph{nets}. One illustration of an netlist is show in Fig. \ref{example1b} a). The netlist can be equivalently described by a directed graph $G(V,E)$ with the set of nodes $V$ and the set of edges $E$. The instances of primitive cells correspond to the nodes $v \in V$ and the nets correspond to edges $e \in E$ of the graph, respectively. As illustrated in Fig. \ref{example1b} b), nets with more than one  sink are described by multiple directed edges between the source primitive cell and the corresponding sink primitive cells. Therefore, each edge $e = (v_i,v_j) \in E$ is directed from $v_i \in V$ to $v_j \in V$.
Between any ordered pair of nodes, there exists at most one directed edge. A sequence of edges $p = (e_1, e_2, e_3, ..., e_n) $ with $e_i = (v_{i-1},v_i)$ is called a path. $p$ is called a cycle, if $v_0 = v_n$. We define the set $K$ as the set of nodes $v \in V$ contained in at least one cycle. Furthermore, we call edge $(v_j,v_i)$ out-edge for the node $v_j$ and in-edge for node $v_i$. The set $E^-(v)$ contains all out-edges and the set $E^+(v)$ all in-edges of node $v$. If $E^+(v)=\{\}$, then $v$ is called primary input node and $v \in PI$, and $PI$ is called the set of primary input nodes. If $E^-(v)=\{\}$, then $v$ is called primary output node and $v \in PO$, and $PO$ denotes the set of primary output nodes.

Finally, as illustrated in Fig. \ref{example1b} b), all nodes and edges will be called \emph{essential}. Moreover, we define all paths $p$ to be {\em critical} if for $p$ holds that the first node $v_0$ is a primary input ($v_0 \in PI$) and the last node $v_n$ belongs to a cycle ($v_n \in K$). All nodes and edges belonging to \emph{critical} paths will be called \emph{critical} as well. As a result, the set of \emph{critical} nodes and edges forms a subset of the set of \emph{essential} nodes and edges.

During the FPGA implementation process, the instances of the primitive cell types \emph{lookup table}, \emph{flip-flop}, as well as \emph{carry logic cells} are first mapped into device-specific basic resource instances $\rho$ of type $\tau_\rho = slice$ which are called \emph{CLB} (configurable logic blocks) slices in case of Xilinx FPGAs.  For example, a single Xilinx CLB slice consists of several lookup tables, flip-flops, carry logic, and multiplexers as well as corresponding nets $E_{CLB}$ connecting them. Typically, a single instance $\rho$ of type $\tau_\rho = slice$ is able to implement several nodes $v$ as well as nets $e$ between these nodes if the required routing resources are present and available. Furthermore, all embedded hard macros, like \emph{multipliers}, \emph{block memory}, \emph{DCMs}, will also be mapped into corresponding device-specific basic resource instances $\rho$ of types $\tau_{\rho} \in \{mult, bram, dcm, \dots \}$. After placing the instances $\rho$, the unrouted nets $E_{con}
 = E \setminus  E_{CLB}$ will be routed. Each routed net $e \in E_{con}$ typically occupies several instances $\rho$ of type programmable interconnect points (PIPs) ($\tau_\rho =  pip$) which are the basic instances of the interconnect network. 

Technically, each instance $\rho$ is configured by a set of configuration bits $B_{\rho} \subset B_{conf}$ with $B_{conf}$ being the union of all configuration bits of the configuration memory of a given FPGA device.
We define the mapping of the instance $\rho$ to the set of configuration bits $B_{\rho}$ by introducing the function $fmap_{\textrm{b}}$ such that
\begin{align}
B_{\rho} = fmap_{\textrm{b}}\left(\rho\right). 
\label{eq:fmap}
\end{align}
In general, this mapping is proprietary and usually unknown to the circuit designer.
In Xilinx FPGAs, the configuration memory is usually organized as a column-based array of equally sized frames which we unify in the set $F$. One frame $f_i \in F$ represents the smallest addressable segments of the configuration memory. For Xilinx Virtex-6 FPGAs, a frame has a height of one clock region and the width of one bit. We denote all bits which belong to one frame $f_i$ as $B_{fi}$.  We define
\begin{align}
B_{\textrm{conf}} = \bigcup_{i = 0}^{N_{\textrm{fr,total}} -1} B_{fi}
\end{align}
with $N_{\textrm{fr,total}}$ being the total number of configuration frames which depends on the size and type of the FPGA device. Now note that the set of used configuration bits $B_\rho$ for implementing an instance $\rho$ might be spread over several configuration frames $f_i$ with
\begin{align}
f_i \in frames\left(B_{\rho}\right) \quad \textrm{and} \quad i \in \{0, 1, ..., N_{\textrm{fr,total}} -1 \},
\end{align}
whereat $frames\left(B_{\rho}\right)$ denotes the set of frames which contain configuration bits of instance $\rho$. For example, the bits which configure a Xilinx CLB slice are distributed over 36 subsequent frames for a Virtex-6 FPGA \cite{rapidsmith}. However, because of the vertical frame structure, one frame also spans over several CLB slices.

In this context, we define $N_{\textrm{fr,used}}$  to be the number of frames containing configuration bits used in a given design. Furthermore, the set of configuration bits $B_{conf}$ can be partitioned into the set of \emph{unused} bits $B_{u}$, a set of \emph{essential} bits $B_{e}$ used in a design and a set of \emph{critical} configuration bits $B_{c}$ as a subset of $B_{e}$ with the distinction that \emph{critical} bits belong to those instances $\rho$ of a circuit that do occur in feedback paths: 
\begin{align}
B_{\textrm{conf}} = B_{\textrm{u}} \cup B_{\textrm{e}} \quad \textrm{and} \quad B_{\textrm{c}} \subseteq B_{\textrm{e}}.
\end{align}
Also, we let $n_{\textrm{e}} = |B_{\textrm{e}}|$, respectively $n_{\textrm{c}} = |B_{\textrm{c}}|$ denote the number of \emph{essential} and \emph{critical} bits.

\section{Definitions w.r.t. Reliability}
\label{sec:notation22}
Traditional scrubbing methods suffer from a long MTTR, since these techniques check and refresh the complete configuration memory frame-by-frame which can take up tens to hundreds of milliseconds \cite{XSEM12}.
Furthermore, by not knowing the criticality of a corrupted bit, the worst case has to be assumed and state-restoring actions, like a global reset or checkpoint restoring, have to be processed as well. This may further increases the MTTR, first by the time $t_{restore}$ needed to perform state-restoring actions and second by the time $t_{lost}$ for reprocessing the lost data. According to \cite{SariP11}, the MTTR can be therefore defined by
\begin{align}
MTTR = MTTD + t_{\textrm{repair}} + t_{\textrm{restore}} + t_{\textrm{lost}}\, ,
\label{eqnMTTR}
\end{align}
where MTTD denotes the \emph{mean time to detect} an SEU and $t_{\textrm{repair}}$ is the time to repair an SEU-corrupted frame. For \emph{essential}, but \emph{non-critical} configuration bits $b \in B_{\textrm{e}} \setminus B_{\textrm{c}}$, we need no state-restoring actions which means that $t_{\textrm{restore}}$ and $t_{\textrm{lost}}$ are zero. For continuous scrubbing methods, the MTTD can be defined as  
\begin{align}
MTTD = t_{\textrm{check}} \cdot \frac{1}{2}\cdot N_{\textrm{fr,used}} ,
\label{eqnMTTD}
\end{align}
where $N_{\textrm{fr,used}}$  is the number of frames which have to be verified and $t_{\textrm{check}}$ is the time to determine if one frame is corrupted or not.

\section{Proposed approach}
\label{sec:background}

Now, in order to decrease the MTTR, we first distinguish and identify automatically \emph{critical} bits from \emph{essential} bits to reduce the number of state-restoring actions. Second, we try to minimize the number of frames $N_{\textrm{fr,used}}$ which have to be verified. 
We achieve the first goal by netlist analysis with subsequent partitioning of primitive cells $v$ and nets $e$ into \emph{critical} and \emph{non-critical} cells and nets (see Section \ref{sec:netlist}). With the help of the Xilinx tool \emph{bitgen}, we are able to determine the corresponding \emph{critical} bits in a given bitfile (see Section \ref{sec:goal}). The great advantage of our method over previous fault-injection approaches like \cite{lee12, Keith05}, is the automatic determination of \emph{critical} bits without requiring any time-consuming bit-wise fault injection and complex verification techniques. Furthermore, after identifying the \emph{critical} cells and nets at logic level, we manipulate the placement and routing to minimize $N_{\textrm{fr,used}}$. Verifying and correcting bits can only be done frame-wise by reading or writing whole frames.
Therefore, the second goal, the reduction of the number of occupied frames $N_{\textrm{fr,used}}$, is achieved by manipulated floorplanning in such a way that a high frame utilization is achieved (see Section \ref{sec:frames}).

\subsection{Analysis of Logic-Level Netlist}
\label{sec:netlist}
The aim of the following netlist analysis is to identify exactly those instances of primitive cells and nets which may lead to a permanent corrupt state of the circuit even if scrubbing would be applied to the SEU-corrupted configuration memory. According to Section \ref{sec:notation}, these instances and nets are obviously located in the netlist in either feedback cycles or on the corresponding input paths to feedback cycles denoted as \emph{critical} instances and nets since a malfunction of these instances or an error in the routing of these nets may lead to erroneous state being trapped in the corresponding cycle.
An SEU occurring in any configuration bit of those components demands a special treatment after correction, like a circuit reset into a valid or fail-save state. Hence, we defined these configuration bits as \emph{critical} configuration bits in Section \ref{sec:notation}. However, such state-restoring actions are not necessary in case an SEU hits a configuration bit belonging to a \emph{non-critical essential} bit.

\subsection{Minimization of the number of used frames \mbox{{${N_{\textrm{fr,used}}}$}}} 
\label{sec:frames}

Typically, traditional scrubbing methods verify all $N_{\textrm{fr,total}}$ frames of the configuration memory, but since only $N_{\textrm{fr,used}}$  configuration frames contain \emph{essential} bits, we suggest to read-back and verify only these $N_{\textrm{fr,used}}$ frames with $N_{\textrm{fr,used}} < N_{\textrm{fr,total}}$. However, if no constraints are used for resource placement and routing, many of the used frames may be low-utilized by \emph{essential} bits and may be scattered all over the FPGA.  

\begin{figure}[!t]
\centering
\includegraphics[scale = 1.2]{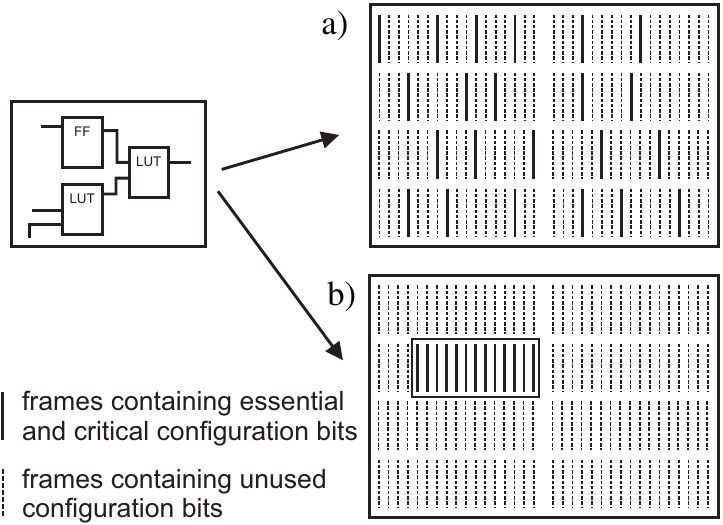}
\caption{Illustration: a) a standard placement and routing of a given circuit and b), our area constrained partitioning and floorplanning. By comparing a) and b), the number of occupied frames is reduced from $N_{\textrm{fr,used}} = 22$ to $N_{\textrm{fr,used}} = 13$ frames. Moreover in b), the cell instances and the routing are aligned to the frame boundaries and placed near to the I/O buffers located in the center of the FPGA \cite{szt14}.}
\label{fig_sim}
\end{figure}

Therefore, by imposing allocation area constraints during the placement process, we first propose to align the used resources of the FPGA device to frame boundaries as illustrated in Fig. \ref{fig_sim} b). To determine the needed number of frames inside a rectangular allocation area, we use the number of utilized resources from the synthesis report. Furthermore, the nearest location to the used I/O pins is chosen for the constrained  area in order to avoid long nets to the I/O pins. Such nets might elsewise generate a huge number of very low utilized additional frames, which must be also checked during scrubbing.  

To avoid net routings leaving the allocated area and generating additional \emph{essential} frames, we constrain also the routing. This can be done by blocking all routing resources outside the allocated area by using so-called \emph{blocker macros}. Such macros can be generated automatically by a floorplanning tool such as \emph{GoAhead} \cite{beck11}. 

The result is a densely placed and routed design where almost all used frames are clustered together. By improving the utilization of frames, also the number of \emph{essential} frames $N_{\textrm{fr,used}}$ is reduced. Another advantage is that the scrubbing may be executed on contiguous regions and without any holes in between. Such holes, which typically result from unconstrained placements, might generate additional scrubbing overhead when using blind scrubbing without readback. If the configuration memory cannot be subsequently written, the scrubber must include an additional configuration header which addresses the frame which should be written next \cite{UG360}.     

\section{Design Flow}
\label{sec:goal}

\begin{figure}[t]
\centering
\includegraphics[width=2.4in]{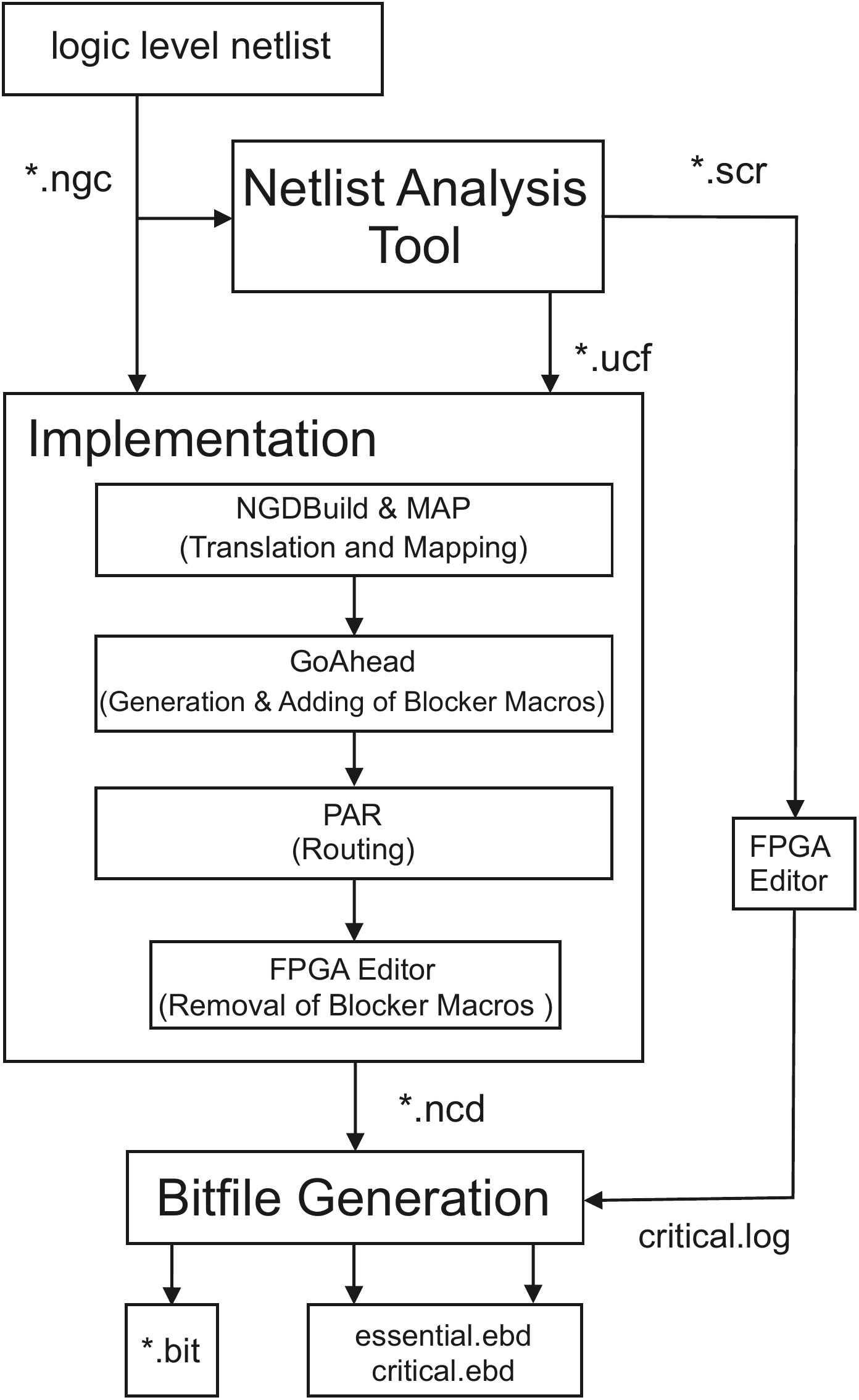}
\caption{Proposed design flow: The partitioning between \emph{essential} and \emph{critical} bits is done in the netlist analysis tool. Furthermore, the floorplanner to minimize the number of occupied frames is controlled by the \emph{ucf} generated by the netlist analysis tool by defining area constraints for the cell instances and by so-called \emph{blocker macros} generated and added by \emph{GoAhead}. The information about the locations of the \emph{essential} and \emph{critical} bits are stored in the \emph{ebd} files \emph{essential.ebd} and \emph{critical.ebd} which are generated by the Xilinx \emph{bitgen} tool.}
\label{fig_designflow}
\end{figure}

Our approach is applicable to many Xilinx SRAM-based FPGAs. The following proposed design flow, as illustrated in Fig. \ref{fig_designflow}, is based on the Xilinx design tool \emph{ISE 14.2}, on our own netlist analysis tool, and on the low-level FPGA design tool \emph{GoAhead} \cite{beck11} which is used for macro generation. Moreover, we use the \emph{Xilinx LogiCORE Soft Error Mitigation (SEM) IP core} \cite{WP395} to implement a scrubbing controller which can detect and correct SEUs in the configuration memory.

The entry point of the proposed design flow is the netlist after synthesis which consists of cell instances and nets given in the Xilinx proprietary \emph{ngc} netlist format. This \emph{ngc} file is converted into the EDIF file format and forwarded to the netlist analysis tool. In addition, the netlist is fed into the implementation stage of the standard Xilinx design flow. 

The netlist analysis tool converts the netlist into the graph $G(V,E)$. From this graph, all \emph{critical} cell instances and nets in the netlist as described in Section \ref{sec:netlist} are extracted. The resulting \emph{critical} cells instances and nets are recorded into a Xilinx FPGA Editor \emph{scr} script file\footnote{\scriptsize{Exceptions are the nets of clock and reset signals which are ignored during our automatic netlist classification but are marked as \emph{critical} nets in the Xilinx FPGA Editor \emph{scr} script file. Furthermore, wildcard characters \emph{*} and \emph{?} are selectively used to tolerate renaming of nets and cell instances during the MAP and PAR optimizations steps.}}. Furthermore, for all cell instances $\rho$, placement constraints are generated and recorded into the Xilinx \emph{User Constraint File} (\emph{ucf}). These placement constraints are arranged such that the number of occupied frames $N_{\textrm{fr,used}}$ is minimized. This is done by aligning the targeted area for implementation of the circuit to the frame boundaries. Moreover, this area is placed next to the fixed I/O buffers in order to avoid long routing lines.  

In the implementation stage, the \emph{ncd} netlist is first translated and mapped to the specific resource instances $\rho$ of the FPGA by using the Xilinx tools \emph{ngdbuild} and \emph{map}, respectively. Afterwards, blocker macros are generated and added to the design via the tool \emph{GoAhead} in order to occupy the routing resources outside the specified area. Without blocking these resources, the Xilinx router might use these resources and generate additional low-utilized frames. After adding blocker macros, the design is placed and routed by the Xilinx \emph{par} tool. Subsequently, the blocker macros are removed again by executing an FPGA-Editor script file, prepared during the generation of the blocker macros.

To identify \emph{essential} configuration bits $B_{\textrm{e}}$ and \emph{critical} configuration bits $B_{\textrm{c}}$ stemming from the corresponding \emph{essential} and \emph{critical} resource instances and nets, we use the Xilinx \emph{bitgen} tool.  A bitfile mask for the \emph{essential} configuration bits $B_{\textrm{e}}$ can be generated directly by \emph{bitgen} using the option parameters \emph{-g essential}. The resulting  \emph{ebd} file which stores for each configuration bit the information if a bit is \emph{essential} or not, can be directly used by the Xilinx SEM IP core. To identify the \emph{critical} bits, we may use the filter function of \emph{bitgen} as described in the Xilinx Application Note 538 \cite{XAPP12}. In this application note, so-called \emph{prioritized essential bits} are introduced which are associated with one or more manually chosen hierarchical modules in the circuit description. To generate an \emph{ebd} file which only consists of \emph{critical} bits from the 
chosen modules, \emph{bitgen} is executed using a filter file. This filter file consists of all resource instances $\rho$ (slices and nets) which belong to the chosen modules. For the proposed approach, we use this filter file to store all \emph{critical} slices and nets which are automatically identified by the method described in Section \ref{sec:netlist}. To identify all \emph{critical} resource instances $\rho$ of type $\tau_{\rho} = slice$ from the \emph{critical} cell instances $v \in V$, we use a script generated by our analysis tool which is executed on the routed netlist by the Xilinx FPGA Editor. 

The results are two different \emph{ebd} files, one for \emph{critical} and one for \emph{essential} bits. These \emph{ebd} files or their containing information may be stored in external and may be used to implement a scrubber which verifies only used frames and finally, initiates state-restoring actions only if a fault on a \emph{critical} bit has been detected.

\section{Configuration Bit Classification through Fault Injection}
\label{sec:faultinj}
\begin{figure}[t]
\centering
\includegraphics[width=3.2in]{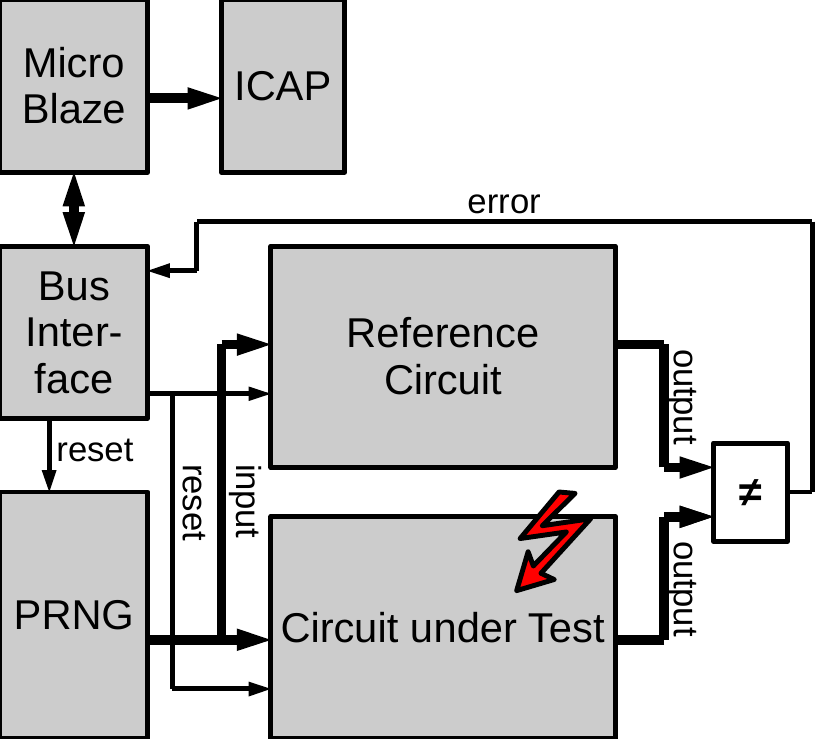}
\caption{The architecture of the FPGA-based fault injection approach. Faults are injected in the configuration of the circuit under test over the ICAP interface. The reference design and the circuit under test are stimulated with identical inputs from the pseudo random number generator (PRNG). A comparator compares the result of both circuits and signals an error to the fault injection software, running on the MicroBlaze.}
\label{fig:faultinj}
\end{figure}

In order to evaluate the results of our systematic approach for configuration bit classification in \emph{essential} and \emph{critical} bits, we implemented a fault injection approach (see Figure \ref{fig:faultinj}). In the past, different fault injection approaches were sucessfully applied to emulate the effects of SEUs in the SRAM cells of the configuration memory \cite{lima2001fault,alderighi2007evaluation,sterpone2007new,mohammadi2012scfit,lee12, Keith05}. The fault injection approach in \cite{Keith05} classifies the configuration bits into \emph{sensitive} and \emph{persistence} bits which is analogous to the \emph{essential} and \emph{critical} classification.  Our fault injection approach is a straight-forward implementation and can be implemented on all modern Xilinx FPGAs without any need of off-chip hardware components.
For the evaluation, we implemented the benchmark circuit in a constrained area of the FPGA and compared the results with a replicated reference circuit implemented in a distinguished area. Faults are injected with the help of a MicroBlaze softcore processor by reading back one configuration frame of the circuit under test over the \emph{internal configuration access port} (ICAP), toggle a single bit, and  a write back to the original location within the configuration memory. It is very important to spatially separate the circuit under test and the reference circuit into different clock regions which corresponds also to different frame addresses. Otherwise, the fault injection might affect also the reference design and, therefore, falsify the result. Both circuits are stimulated with the same input pattern, produced by a pseudo random number generator.

In order to determine the \emph{essential} and \emph{critical} bits with fault injection, every configuration bit of the constrained area on which the circuit under test is implemented must be toggled with subsequent evaluation of the impact. This is done at follows: First, the circuit under test and the reference design is reset. The fault is injected in the corresponding configuration bit by reading back the corresponding frame, toggle the bit, and write it back into the configuration memory. After that, the circuit under test and the reference circuit process input data stemming from the pseudo random number generator for a specified time. The outputs of both circuits are compared and if a mismatch is discovered during the processing, the corresponding configuration bit is marked as \emph{essential}. After that, the error-free frame is recovered in the configuration memory and the testing sequence is replayed without reseting the circuits by reseting only the pseudo random number generator.  If an error occurs even on the now error-free configuration, the configuration bit to test is additionally marked as \emph{critical}. Finally, both circuits are reset and the next configuration bit is evaluated. This is repeated until all configuration bits in the constrained area are tested. The corresponding test program which implements the described procedure is implemented on the MicroBlaze processor.    

\section{Experimental Results}
\label{sec:analysis}

The design flow described in Section \ref{sec:goal} was applied to a subset of 20 benchmark circuits from the MCNC suit obtained at \cite{bench}. Only circuits with flip-flops were considered. Additionally, we choose a \emph{LMS equalizer}, a \emph{floating-point unit} (FPU), an \emph{AES 128-bit cipher} implementation and a \emph{(204,188)-Reed-Solomon Decoder} from Opencores.org \cite{opencores}. For our experiments, the circuits were implemented for the Virtex-6 \emph{XC6VLX240T} FPGA. In the following, Subsection \ref{sec:floor2} shows our results regarding the floorplanning and Subsection \ref{sec:mttr2} shows our results regarding the MTTR.

\subsection{Reducing the number of used frames \mbox{{${N_{\textrm{fr,used}}}$}}}
\label{sec:floor2}

In Table \ref{tab:PPer}, the benchmark designs are characterized in terms of utilized LUTs, flip-flops, I/O buffers and the number of frames $N_{\textrm{fr,FF}}$ which contain the configuration bits of the utilized flip-flops\footnote{\scriptsize{There are no \emph{embedded multipliers} or BRAMs instantiated in the benchmark designs. While our autmatic netlist analysis approach may apply to \emph{embedded multipliers}, it does not apply to BRAMs since the associated content of the SRAM cells is altered during circuit operation and cannot be scrubbed.}}. Furthermore, the table shows the number of identified \emph{essential} bits $n_{\textrm{e}}$, the number of identified \emph{critical} bits $n_{\textrm{c}}$, the ratio $n_{\textrm{c}} / n_{\textrm{e}}$ for the case that no additional placement and routing constraints are used.
In Table \ref{tab:PPer2}, the numbers of occupied frames $N_{\textrm{fr,used}}$ are depicted for the case a) that no placement and no routing constraints are used, for the case b) that placement constraints and no routing constraints are used, and for the case c) that placement and routing constraints are used. The last two columns show the reductions of the number of used frames $N_{\textrm{fr,used}}$ when case c) is compared to case a) and case b), respectively.

The approach in \cite{SariP11} uses only placement constraints which corresponds to our case b). Therefore, the comparison between case b) and c) in the last column of Table \ref{tab:PPer2} depicts our improvement over the approach in \cite{SariP11} by using additional routing constraints.

\begin{table*}[bth]
\caption{Benchmark circuits with used LUTs, flip-flops, IO Buffers, the number of frames $N_{\textrm{fr,FF}}$ containing configuration bits of user logic flip-flops, the number of \emph{essential} and \emph{critical} bits.}
\centering
\small
\resizebox{\linewidth}{!}{
\begin{tabular}{ |l|| c | c | c| c | c | c | c | c|c|}
\hline
\textbf{Circuit}  & \textbf{LUTs} & \textbf{FFs} & \textbf{IO Buffers} & {$N_{\textrm{fr,FF}}$} & {$n_{\textrm{e}}$} & {$n_{\textrm{c}}$} & {$n_{\textrm{c}}$} / {$n_{\textrm{e}}$} \\ \hline \hline 
\multicolumn{8}{|c|}{\textbf{\emph{MCNC benchmark circuits}}} \\ \hline \hline
\textbf{\emph{bigkey}  }  & 536    & 224 & 426 &  33     & 279584         & 250737       & 0.897 \\ \hline
\textbf{\emph{diffeq}}    & 555    & 248 & 31  &  15     & 205813         & 205813       & 1      \\ \hline
\textbf{\emph{elliptic}}  & 109    & 63  & 5   &   8     & 112832         & 112832       & 1      \\ \hline
\textbf{\emph{frisc}   }  & 1889   & 822 & 136 &  23     & 529562         & 527779       & 0.997 \\ \hline
\textbf{\emph{s38417}}    & 2188   & 1263& 135 &  37     & 502052         & 411022       & 0.819 \\ \hline
\textbf{\emph{s38584.1}}  & 1693   & 1140& 342 &  41     & 465790         & 386888       & 0.831 \\ \hline
\textbf{\emph{tseng}}     & 478    & 221 & 174 &  19     & 216002         & 196154       & 0.908 \\ \hline \hline
\multicolumn{8}{|c|}{\textbf{\emph{Opencores benchmark circuits}}} \\ \hline \hline
\textbf{\emph{LMS equalizer}} & 231    & 206   & 34  &   96   & 156264         & 156264            & 1      \\ \hline
\textbf{\emph{FPU}} & 8007    & 553   & 110  &   171   & 1713480         & 1680509            & 0.981     \\ \hline
\textbf{\emph{AES 128-bit }} & 10450    & 10769   & 389  &   316   & 2387020         & 2372907   & 0.994           \\ \hline
\begin{tabular}{@{}l@{}}\textbf{\emph{(204,188)-RS}}\\ \textbf{\emph{decoder}}\end{tabular}  & 3808    & 2735   & 21  &   183   & 883437         & 858757            & 0.972     \\ \hline
\end{tabular}}
\label{tab:PPer}

\end{table*}

\begin{table*}[bth]
\caption{Benchmark circuits from Table \ref{tab:PPer} with the number of occupied frames without (a)) and with placement constraints (b)) and with placement and routing constraints (c)). The last two columns show the reductions of the number of used frames $N_{\textrm{fr,used}}$ when case c) is compared to case a) and case b), respectively.}
\centering
\small
\resizebox{\linewidth}{!}{
\begin{tabular}{ |l|| c|c| c |c| c|}
\hline
\textbf{Circuit}  & {\begin{tabular}{@{}c@{}}$N_{\textrm{fr,used}}$ \\ for a) \end{tabular}} & {\begin{tabular}{@{}c@{}}$N_{\textrm{fr,used}}$ \\ for b) \end{tabular}}& {\begin{tabular}{@{}c@{}}$N_{\textrm{fr,used}}$ \\ for c)  \end{tabular}} & {\begin{tabular}{@{}c@{}}$\Delta N_{\textrm{fr,used}}$ in (\%) \\ comparing c) and a)  \end{tabular}} & {\begin{tabular}{@{}c@{}}$\Delta N_{\textrm{fr,used}}$ in (\%) \\ comparing c) and b)  \end{tabular}} \\ \hline \hline 
\multicolumn{6}{|c|}{\textbf{\emph{MCNC benchmark circuits}}} \\ \hline \hline
\textbf{\emph{bigkey}  }  &  4201   & 2393            & 1891  & 43 & 21.0 \\ \hline
\textbf{\emph{diffeq}}    &  1332   & 602             & 602   & 54.8 & 0.0  \\ \hline
\textbf{\emph{elliptic}}  &   912    & 395             & 395  & 56.7 & 0.0  \\ \hline
\textbf{\emph{frisc}   }  &  1960   & 1292            & 1050  & 34.1 & 18.7 \\ \hline
\textbf{\emph{s38417}}    & 2027   & 1389            & 1389   & 31.5 & 0.0  \\ \hline
\textbf{\emph{s38584.1}}  & 3192   & 2270            & 1863   & 28.9 & 17.9 \\ \hline
\textbf{\emph{tseng}}     & 1924   & 1298            & 1054   & 32.5 & 18.8 \\ \hline \hline
\multicolumn{6}{|c|}{\textbf{\emph{Opencores benchmark circuits}}} \\ \hline \hline
\textbf{\emph{LMS equalizer}} &  2051   & 1563             & 1120   & 23.8  & 28.3  \\ \hline
\textbf{\emph{FPU}}  &  4117   & 3520             & 3280   & 11.5 & 6.8  \\ \hline
\textbf{\emph{AES 128-bit }}  &  6822   & 6657       & 6335   & 2.4 & 4.8  \\ \hline
\begin{tabular}{@{}l@{}}\textbf{\emph{(204,188)-RS}}\\ \textbf{\emph{decoder}}\end{tabular}  &  4610   & 4073       & 3761   & 11.6 & 7.7  \\ \hline
\end{tabular} }
\label{tab:PPer2}

\end{table*}
It can be seen that by comparing case c) to case a), the number of occupied frames can be reduced by up to 59\% by using placement and routing constraints and by comparing c) to b), the number of occupied frames can be reduced by up to 23.7\% just by using additional routing constraints. 
These gains directly affect the achievable MTTD as well, since according to Eq. (\ref{eqnMTTD}), the MTTD is linearly dependent on the number of occupied frames. Nevertheless, the achievable gains may vary considerably from design to design and depends on the distribution of the used I/Os.
Furthermore, our experiments show that either no (0\%) (complete combinatorial circuit) or between 81.9\% to 100\% of the \emph{essential} bits are also \emph{critical} bits in the analyzed circuits. But nevertheless, in case of the circuit \emph{s38417}, if an SEU occurs on a used configuration bit, there is still the probability of 18.1\% that no state-restoring action has to be carried out after scrubbing.

\begin{figure*}[!t]
\centering
\includegraphics[scale=0.8]{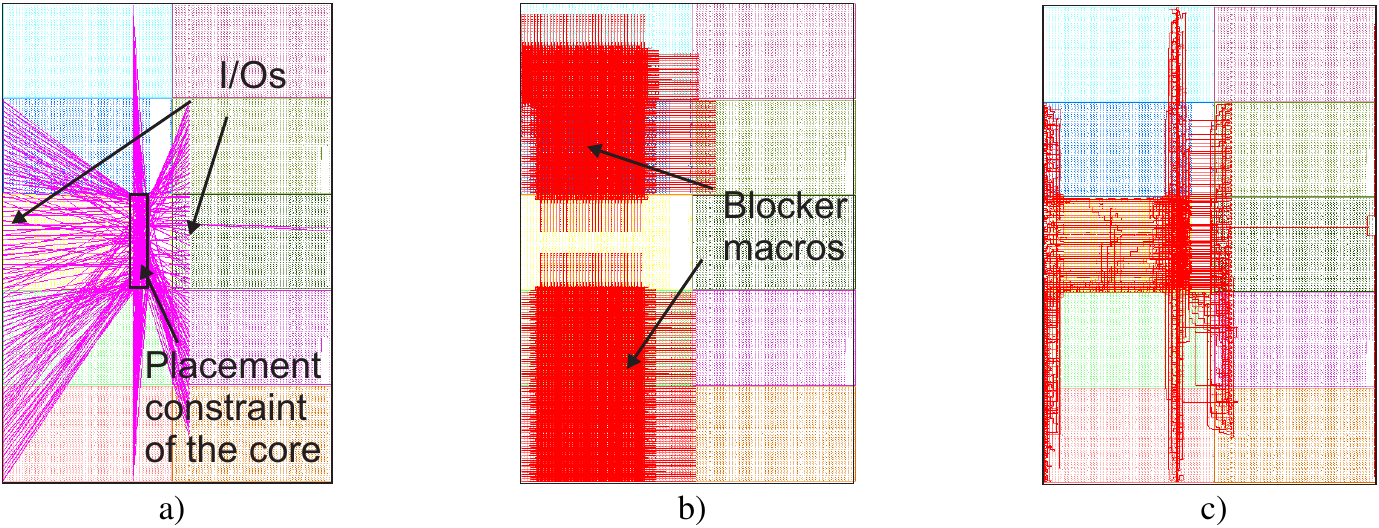}
\caption{Illustration of our floorplanning approach applied to one of our benchmark circuits. In a), the unrouted, but placed design is shown. The placement area is highlighted by the black box. In b), the blocker macros are shown which impose the routing constraints. In c), the placed and routed design is shown.}
\label{designexample}
\end{figure*}

Fig. \ref{designexample} illustrates our floorplanning strategy for one of our benchmark designs. Fig. \ref{designexample}a) shows the placed, but unrouted design. The placement constraint is highlighted by the black frame which is aligned to the frames of the middle left clock region. It can be seen that many wires have to be routed from the I/O buffers of the left side to the I/O buffers in the middle. Therefore, as shown in Fig. \ref{designexample}b), blocker macros are generated with the tool \emph{GoAhead} to force the wires to be routed through the middle left clock region. This is done to raise the frame utilization{\footnote{\scriptsize{The normalized frame utilization of one frame is defined the ratio of the number of \emph{essential} bits to the total number of configurations bits in a frame.}}} in this area and to minimize the additional frames occupied by routing resources. Fig. \ref{designexample}c) shows the final optimized design without the removed blocker macros, in which all routing wires are mainly routed through the desired clock region. Note that placement and routing constraints have also an effect on the timing. The effect depends on the size of the constrained area, the circuit and the desired clock frequency. The decrease of the maximum clock frequency is usually below 10\% (see also \cite{KT10}). 

\begin{figure*}[ht]
\centering
\includegraphics[scale = 1.00]{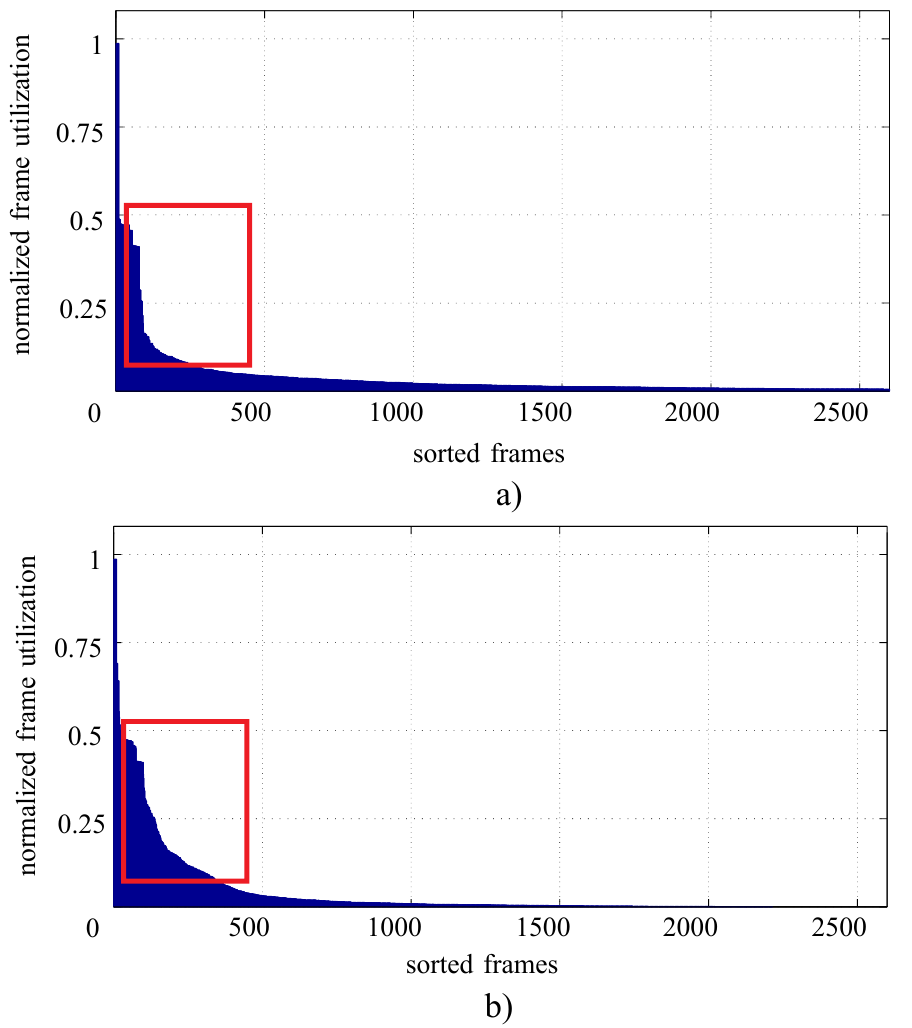}
\caption{The diagrams show the frame utilization of the benchmark circuit \emph{bigkey} for the case that no special placement and routing techniques are used (a)) and for the case that our proposed placement and routing constraints are used (b)). The frames are sorted according to their frame utilization in terms of the number of \emph{essential} bits in descending order. Due to the placement and routing constraints, the number of occupied frames $N_{\textrm{fr,used}}$ is reduced from 4201 to 1891 and, therefore, the frame utilization shown in b) is higher.}
\label{ut}
\end{figure*}

Fig. \ref{ut} shows two diagrams of the frame utilization for the benchmark design \emph{bigkey}. Fig. \ref{ut}a) illustrates the frame utilization without using any special placement and routing constraints, and Fig. \ref{ut}b) illustrates the frame utilization with placement and routing constraints. The frames in both diagrams are sorted in descending order from left to right regarding to the frame utilization with \emph{essential} bits. It can be seen that the \emph{essential} bits are grouped in less frames and that, in general, the utilization of the frames containing \emph{essential} bits is higher when adding our placement and routing constraints.

\subsection{Comparison to Fault Injection Approach}

The netlist configuration bit classification method is compared to the fault injection approach, described in Section \ref{sec:faultinj}. Hereby, we selected the three largest MCNC benchmark circuits (\emph{frisc}, \emph{s38417}, and \emph{s38584.1}) for comparison (see Table \ref{tab:PPerFI}). We count only the  essential bits $n_e$ and critical bits $n_c$ for the constrained area in which the circuit under test resides. Therefore, the number of determined essential and critical bits of the netlist method $n_e$ and $n_c$ differ to the values in Table \ref{tab:PPer}. The reason for the smaller number of essential bits in Table \ref{tab:PPerFI} might be the missing long connections to the I/O cells and the more dense packing.

\begin{table*}[bth]
\caption{The results of the classification by fault injection method of three selected benchmark circuits. First, the numbers of essential bits $n_{\textrm{e}}$ and critical bits $n_{\textrm{c}}$ detected by the netlist and fault injection method are shown. Furthermore, the ratios between essential and critical bits detected by the fault injection and the netlist method are presented. }
\centering
\small
\resizebox{\linewidth}{!}{
\begin{tabular}{ |l|| c | c | c| c | c |c|c|}
\hline
\textbf{Circuit}  & {netlist} & {netlist} & {fault injection}& fault injection  & fault injection vs.  & fault injection vs.\\ 
 & method {$n_{\textrm{e}}$}& method {$n_{\textrm{c}}$}  & method {$n_{\textrm{e}}$} & method {$n_{\textrm{c}}$} & netlist method  {$n_{\textrm{e}}$} & netlist method  {$n_{\textrm{c}}$}\\\hline \hline 
\multicolumn{7}{|c|}{\textbf{\emph{MCNC benchmark circuits}}} \\ \hline \hline
\textbf{\emph{frisc}   }  & 338218  & 337542 & 30977  & 5215 &  9 \%  &1.5 \% \\ \hline
\textbf{\emph{s38417}}    & 311584  & 255810  & 17677 & 457 &  6 \% &  0.2 \% \\ \hline
\textbf{\emph{s38584.1}}  & 313059  &261091 & 160008& 0 &  51 \%   &  0 \% \\ \hline
\end{tabular}}
\label{tab:PPerFI}
\end{table*}

The constrained area has the height of one clock region and the width of  up to 8 CLB columns. This results in up to 746,496 configuration bits which have to be evaluated. For each configuration bit, the test sequence consisting of 1 million input vectors, generated by the pseudo random number generator has to be applied twice (see Section \ref{sec:faultinj}). The overall test time for one core is up to 20 hours. 

The values in Table \ref{tab:PPerFI} show that only a subset of thenetlist method's \emph{essential} bits is found by the fault injection method. The reason is that the \emph{essential} bits determined by Xilinx \emph{Bitgen} is a worst case estimation. For example, if only two inputs of a LUT are used, most bits are not used whereas \emph{Bitgen} will classify all configuration bits belonging to this LUT as \emph{essential}. Moreover, due to circuit inherent redundancy and false paths, a lot of errors are masked and have, therefore, no impact on the outputs. However, the one million input test vectors are far away to be enough to discover all possible configuration faults which lead to a circuit failure.
Moreover, the number of found \emph{critical} bits is much lesser than the number of found \emph{essential} bits. Here, only a small fraction of the determined bits from the netlist method is identified. To understand this behavior, we have to emphasize that the netlist method categorize all bits belonging to feedback paths as \emph{critical} and \emph{essential}. The experiment for determining the \emph{essential} bits with the fault injection method shows, that only a small part of \emph{essential} bits which were found by the netlist method has the ability to corrupt the output due to the above mentioned masking effects. However, we applied one million test vectors whichmeans that the fault has plenty of time to corrupt at least one single bit in one output vector, but during the second run which identifies the \emph{critical} bits, only the falsified states in the registers at the end of the first run has the ability to corrupt the outputs. It could also be the case that the state in the registers during switching from the first to second run was correct and an error on the outputs occurred much earlier during the first run. Even if the state is actually wrong, the falsified state might has no effect on the following states due to the above mentioned masking effects. This shows us that the results of the fault injection methods, especially for the \emph{critical} bits,  have to be handled with care. 

On the other hand, all discovered \emph{critical} and \emph{essential} bits by fault injection are also included in the set of \emph{essential} and \emph{critical} bits of the netlist method. Therefore, the netlist method is surely an overestimation, but by using this method, we can be sure that we will find any configuration fault which leads to a failure. By avoiding the time-consuming tests of the fault injection method which must be applied after each new place and route iteration, the proposed netlist analysis  is easy to use and produces very fast the corresponding result.

\subsection{MTTR Analysis}
\label{sec:mttr2}

In order to calculate the MTTR, we assume a scrubbing system with checkpointing as described in \cite{SariP11}. Additionally, we assume that this system uses the SEM IP core \cite{XSEM12}. All calculations rely on the duration times given in \cite{XSEM12}. 
The SEM IP core can be directly used to support our classification approach of \emph{critical} bits. However, the original SEM do not support selective scrubbing approach. To also support this feature, the core has to be modified. The original SEM IP core requires 459 LUTs and 383 flip-flops.Our experimental scrubber core with our SEM IP Core modifications requires 2706 LUTs and 2176 flip-flops.

\begin{figure*}[ht]
\centering
\includegraphics[scale = 0.65]{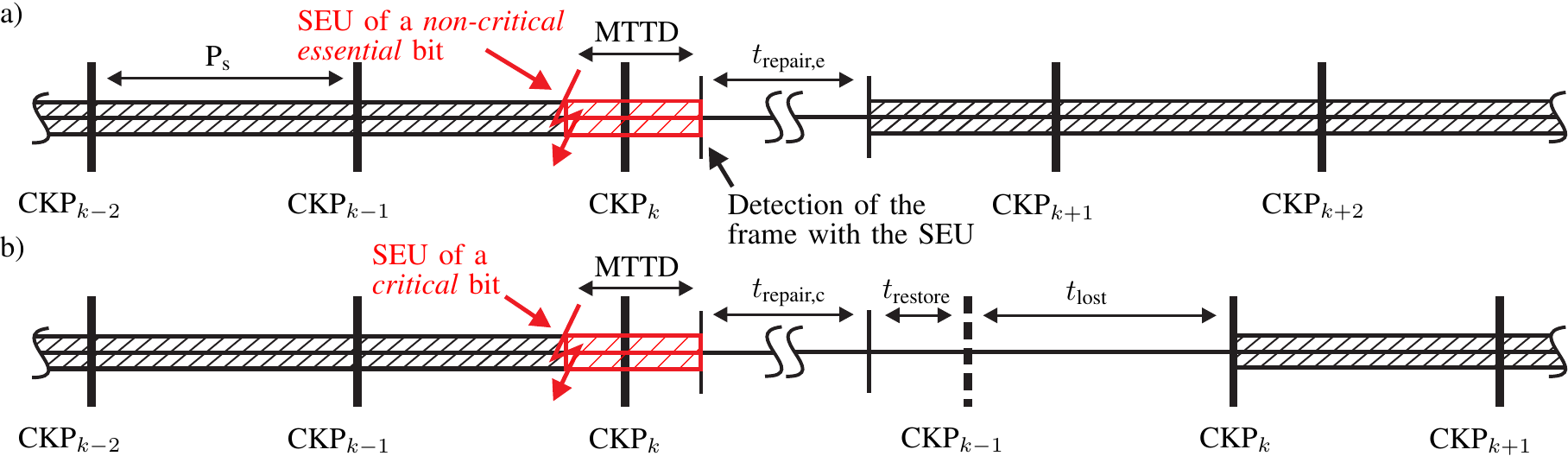}
\caption{Illustration of the adopted checkpointing system, in which the checkpoints $\textrm{CKP}_k$ with the time index $k$ are created in time intervals equal to the scan duration $\textrm{P}_\textrm{s}$. In a), a \emph{non-critical essential} bit is corrupted by an SEU. The corresponding corrupted frame is detected on average after the MTTD. Afterwards, the corrupted \emph{essential} bit is detected within the frame and repaired, which in total takes the time $t_{\textrm{repair,e}}$. In b), a \emph{critical} bit is corrupted by an SEU. In this case, the system has to be rolled back to the checkpoint $\textrm{CKP}_{k-1}$ after the corrupted frame has been detected (MTTD) and repaired ($t_{\textrm{repair,c}}$).
The rollback to the checkpoint $\textrm{CKP}_{k-1}$ takes the time $t_{\textrm{restore}}$. Furthermore, it takes the time $t_{\textrm{lost}} = \textrm{P}_\textrm{s}$ until the state of checkpoint $\textrm{CKP}_{k}$ has been reached again \cite{szt14}.}
\label{ut2}
\end{figure*}

As illustrated in Fig. \ref{ut2}, we assume that the occupied frames of the configuration memory are continuously scanned for SEUs and checkpoints $\textrm{CKP}_k$ with the time index $k$ are created in intervals equal to the scan duration $\textrm{P}_\textrm{s}$ with $\textrm{P}_\textrm{s} = N_{\textrm{fr,used}} \cdot t_{\textrm{check}}$. If a \emph{non-critical} \emph{essential} bit is corrupted, as illustrated in Fig. \ref{ut2} a), the scrubbing system needs on averaged the time MTTD to detect the corrupted frame and repairs the corresponding bit with no further action, afterwards. The corresponding repair time $t_{\textrm{repair,e}}$ of an \emph{essential} but \emph{non-critical} bit includes the time to determine the corrupted bit within the corrupted frame and the time to write the repaired frame data into the configuration memory. However, if a \emph{critical} bit is corrupted, as illustrated in Fig. \ref{ut2} b), the system state is rolled back to a valid checkpoint after the detection and correction 
of the corrupted bit. The system is rolled back to the checkpoint $\textrm{CKP}_{k-1}$ under the assumption that the corrupted \emph{critical} bit has been detected after the checkpoint $\textrm{CKP}_{k}$. Therefore, at least the two past checkpoints $\textrm{CKP}_{k}$ and $\textrm{CKP}_{k-1}$ have to be saved \cite{SariP11}. The repair time $t_{\textrm{repair,c}}$ of a \emph{critical} bit is in general greater than $t_{\textrm{repair,e}}$ since $t_{\textrm{repair,c}}$ further includes the classification time to check if the corrupted bit is indeed a \emph{critical} configuration bit, e.g., by evaluating an \emph{ebd} file stored in external memory. In a frame, which contains \emph{essential} bits, every SEU has to be corrected even if the SEU occurred on an unused bit since the frames are corrected by an evaluation of the ECC parity bits which can only correct one error per frame. Therefore, no accumulation of SEUs is allowed\footnote{\scriptsize{Our approach may be also applicable to multiple bit upsets as long as there is at most one bit-flip in one frame of the configuration memory.}}. However, SEUs on unused bits do not contribute to the MTTR because the 
implemented circuit is not damaged. 
After a detected SEU on an \emph{essential} bit, the criticality has to be determined (classification time). This step is always performed, however it contributes only to the MTTR if the SEU occurred on a \emph{critical} bit, because after the correction of an \emph{essential} \emph{non-critical} bit, the circuit is again fully functional\footnote{\scriptsize{The \emph{ebd} file with the \emph{essential} bits is just needed to determine the frames which have to be scrubbed. Only the \emph{ebd} with the \emph{critical} bits is used for classification.}}. In order to set the registers into the state of the checkpoint $\textrm{CKP}_{k-1}$, the configuration bits of all used flip-flops have to be set to the stored register values. Therefore, $N_{\textrm{fr,FF}}$ frames have to be re-written after $t_{\textrm{repair,c}}$.

For our analysis, we assume that re-writing one frame to restore the state of a checkpoint requires the duration of two times $t_{\textrm{check}}$ which on the one hand includes the overhead to write the frame with the mask bit to restore the corresponding flip-flops values and which on the other hand also includes the overhead to update the ECC parity bits. Therefore, we define $t_{\textrm{restore}} = 2 \cdot t_{\textrm{check}} \cdot N_{\textrm{fr,FF}}$. Furthermore, the rollback to the checkpoint $\textrm{CKP}_{k-1}$ induce $t_{\textrm{lost}} = N_{\textrm{fr,used}} \cdot t_{\textrm{check}}$ which equals the time to reach the state of $\textrm{CKP}_{k}$. In case of a corrupted \emph{essential} bit that is \emph{non-critical}, we set $t_{\textrm{restore}}$ and $t_{\textrm{lost}}$ to zero, since we assume that the system tolerates errors induced by a corrupted \emph{essential} but \emph{non-critical} bit and recovers after correction on its own.
For our calculations of the MTTR, we propose the following equation:
\noindent
\begin{align}
MT&TR =  \left(\frac{n_{\textrm{e}}-n_{\textrm{c}}}{n_{\textrm{e}}}\right) \cdot \left(MTTD + t_{\textrm{repair,e}}\right) \notag \\
&+ \left(\frac{n_{\textrm{c}}}{n_{\textrm{e}}}\right) \cdot \left(MTTD + t_{\textrm{repair,c}}  + t_{\textrm{restore}} + t_{\textrm{lost}} \right).  
\label{eq:MTTR2}  
\end{align} 
According to the product guide of the SEM IP core \cite{XSEM12}, we set the repair time of \emph{essential} bits $t_{\textrm{repair,e}} = 490  \, \mu s $ and the repair time of \emph{critical} bits $t_{\textrm{repair,c}} = 1100  \, \mu s $ which additionally includes $ 610 \, \mu s$ for the classification. Furthermore, we obtain $t_{\textrm{check}} = 810 \, ns$ which is determined by the frame size of 2592 bits, the ICAP interface word size of 32 bit and the ICAP frequency of 100 MHz.

In Table \ref{tab:PPer3}, we compare the MTTR of four different types of scrubbing controllers: a) \emph{scrubbing without any classification}, b) \emph{scrubbing with classification of unused and essential bits}, c) \emph{scrubbing with classification of used, essential, and critical bits}, and d) \emph{scrubbing with classification of used, essential, and critical bits and with placement and routing constraints}. For the type a) the scrubbing controller continuously scans the whole configuration memory consisting of $N_{\textrm{fr,total}} = 22261$ frames and initiates a reset with rollback after each error correction. Since we use a rather huge FPGA device for our small benchmark circuits, the application of this scrubber is not reasonable for our evaluation, but the MTTR is given for the sake of completeness\footnote{The main part of these huge MTTRs are the terms MTTD and $t_{lost}$ which corresponds directly to $N_{fr,total}$ and needs $MTTD + t_{lost} = 27137 \mu s$. For the smallest available Virtex-6 FPGA, the XC6VLX75T, this time is $9234 \mu s$.}. A scrubbing controller of type b), scans only the occupied frames, but also initiates a reset with rollback even if the corrupted bit is \emph{non-critical}.  The scrubbing controller of type c) only 
scans the occupied frames and only initiates a reset in case a \emph{critical} bit is corrupted. We finally suggest and compare a scrubbing controller of type d) which also uses classification but just scans a reduced set of frames due to the placement and routing constraints. In this analysis, the scrubbing controllers of type a) and b) treat every corrupted configuration bit as a corrupted \emph{critical} configuration bit. Apparently, our proposed approach clearly outperforms the scrubbing controller types a) and b). If it is possible to reduce the number of occupied frames by using placement and routing constraints, then d) outperforms c) as well. In comparison to type b), which can be seen as a standard scrubbing controller, the savings regarding to the MTTR are up to 48.5\% for our benchmark circuits.

Table \ref{tab:PPer3} brings out that the savings regarding the MTTR heavily depends on the circuit. Furthermore, by analyzing complex circuits and circuits with a high utilization of the FPGA device, most of the \emph{essential} bits will indeed be identified as \emph{critical} bits. This is especially true for circuits which are not datapath-intensive.   

\begin{table*}[th]
\small
\caption{Comparison of the MTTR for the considered benchmark circuits using four different types of scrubbing controllers: a) \emph{scrubbing without any classification}, b) \emph{scrubbing with classification of unused and essential bits}, c) \emph{scrubbing with classification of used, essential, and critical bits}, and d) \emph{scrubbing with classification of used, essential, and critical bits and with placement and routing constraints}. The last three columns show the savings regarding the MTTR when type d) is compared to type a), type b), and type c).}
\centering
\resizebox{\linewidth}{!}{
\begin{tabular}{ |l|| c | c | c|  c|c |c|c|}
\hline
& \multicolumn{4}{|c|}{MTTR in [$\mu$s]} & \multicolumn{3}{|c|}{$\Delta$ MTTR in (\%) comparing} \\
\textbf{{Circuit}} &  for a)  &  for b)  &  for c)  &  for d) &  d) and a)  &  d) and b)   &   d) and c) \\ \hline \hline
\multicolumn{8}{|c|}{\textbf{\emph{MCNC benchmark circuits}}} \\ \hline \hline
\textbf{\emph{bigkey}}  & 27590       & 6258     &    5838         & 3224  &     88.3  & 48.5 & 44.8  \\ \hline
\textbf{\emph{diffeq}}  & 27561       & 2742     &    2741         & 1855  &     93.3  & 32.3 & 32.3  \\ \hline
\textbf{\emph{elliptic}}& 27550       & 2221     &    2221         & 1592  &     94.2  & 28.3  & 28.3  \\ \hline
\textbf{\emph{frisc}}   & 27574       & 3519     &    3511         & 2408  &     91.3  & 31.6 & 31.4  \\ \hline
\textbf{\emph{s38417}}  & 27597       & 3628     &    3203         & 2522  &     90.9  & 30.4 & 21.3  \\ \hline
\textbf{\emph{s38584.1}}& 27603       & 5045     &    4492         & 3059  &     88.9  & 39.3 & 31.9  \\ \hline
\textbf{\emph{tseng}}   & 27567       & 3468     &    3266         & 2274  &     91.8  & 34.4 & 30.4  \\ \hline \hline
\multicolumn{8}{|c|}{\textbf{\emph{Opencores benchmark circuits}}} \\ \hline \hline
\textbf{\emph{LMS equalizer}} & 27693 & 3748 & 3748 & 2616 & 90.6 & 30.2 & 30.2 \\ \hline
\textbf{\emph{FPU}} & 27814 & 5769 & 6379 & 5294 & 81.0 & 17.0 & 15.9 \\ \hline
\textbf{\emph{AES 128-bit}} & 28049 & 9901 & 9861 & 9272 & 66.9 & 6.4 & 6.0 \\ \hline
\begin{tabular}{@{}l@{}}\textbf{\emph{(204,188)-RS}}\\ \textbf{\emph{decoder}}\end{tabular}  & 27834 & 6998 & 6868 & 5856 & 79.0 & 16.3 & 14.7 \\ \hline

\end{tabular}}
\label{tab:PPer3}
\end{table*}

\section{Conclusions and Future Work}
\label{sec:conclusion}

In this work, we present two new methods to enhance common scrubbing techniques for SEU mitigation: a) we apply a netlist analysis and the Xilinx tool \emph{bitgen} to identify and distinguish \emph{essential} and \emph{critical} bits in the configuration memory and b) we use placement and routing constraints to align a given design to the frame boundaries in order to reduce the number of occupied frames that are affected by scrubbing. We integrated these two methods into one design flow, which is fully automated by script files. A comparison to an implemented alternative classification approach using fault injection is given which shows that through the proposed netlist analysis, time-consuming fault injection methods can be avoided to identify \emph{essential} and \emph{critical} configuration bits. Furthermore, due to differentiation between \emph{unused}, \emph{essential}, and \emph{critical} bits, we can efficiently decide if an SEU in the configuration memory has to be corrected at all or if an SEU demands a reset after scrubbing.
As was shown by experiments, the introduction of placement and routing constraints may not only help to minimize the number of occupied frames, but also may reduce the MTTR for specific designs by up to 48.5\% in comparison to a standard scrubbing controller.

The proposed design flow and the two methods open many possibilities for future work. As seen in the experimental results, the number of \emph{critical} bits dominates the \emph{essential} \emph{non-critical} bits on the most real world examples. However, error masking through exploitation of redundancy mechanisms and subsequently voting are not considered in our netlist analysis. This is also indicated by the comparison to the fault injection method which identify much lesser bits as \emph{essential} and \emph{critical}. By the sophisticated usage of methods such as DMR and TMR inside feedback loops, we are able to establish an error barrier to mask out an error by voting before it can affect the state of the circuit permanently. This means that former \emph{critical} bits of these feedback loops might be transferred into \emph{non-critical} bits which lowers the overall number of \emph{critical} bits in a design. This might be reached by using the STAR framework \cite{sterpone2005new}. For such designs we predict a higher benefit by using our proposed methodology to reduce the MTTR.

Furthermore, we did not consider any circuits which utilize BRAMs or which implements state-driven protocols to communicate with external peripherals, e.g., external memory. Both the interfaces to BRAMs and
the interfaces to external peripherals may lead to feedback cycles which are not handled by our methodology so far. 
The consideration of memories will therefore extend the scope of our work even further.

For newer Xilinx FPGA families, like 7-series and Ultrascales, the Xilinx design tool suite \emph{Vivado} has to be used. Our approach relies on the tool \emph{GoAhead} \cite{beck11} which uses the XDL interface of the \emph{ISE} design tools. However, a new version is under development which utilizes the TCL interface of Vivado.

\section*{Acknowledgments}

The work has been partially supported by EFRE funding
from the Bavarian Ministry of Economic Affairs (Bayerisches
Staatsministerium f\"ur Wirtschaft, Infrastruktur, Verkehr und
Technologie) as a part of the ``ESI Application Center'' project.


\bibliography{SEU_RAW2014_refs}

\end{document}